\documentclass[11pt,a4paper]{article}

\usepackage[margin=1in]{geometry}
\usepackage{amsmath,amssymb,amsthm}
\usepackage{mathtools}
\usepackage{bm}
\usepackage{graphicx}
\usepackage{booktabs}
\usepackage{hyperref}
\usepackage{cite}
\usepackage{enumitem}
\usepackage{physics}
\usepackage{xcolor}

\hypersetup{
	colorlinks=true,
	citecolor=blue,
	linkcolor=blue,
	urlcolor=blue
}

\theoremstyle{definition}

\theoremstyle{remark}
\newtheorem{remark}{Remark}[section]

\title{\bfseries Non-Stationary Decoherence in Superconducting Qubits:\\[3pt]
	Memory Multi-Fractional Brownian Motion and a Time-Dependent\\[3pt]
	Quantum Brownian Motion Extension}

\author{}
\date{}

\begin{document}
	\maketitle
\author{Mahboob Ul Haq}
\date{\today}	
	\begin{abstract}
		Building upon our prior work~\cite{Haq2025}, we present a unified stochastic drift model (SdM) for superconducting charge qubits based on memory multi-fractional Brownian motion (mmfBm). The classical sector employs a time-dependent Hurst exponent $H(t)$ and adaptive memory kernel $K(t,s)$, capturing non-stationary $1/f^\beta$ noise and long-range temporal correlations inaccessible to conventional models. The quantum extension is formulated via a time-dependent Caldeira--Leggett environment with spectral density $J(\omega; t) = \eta(t)\,\omega_c^{1-s(t)}\omega^{s(t)}e^{-\omega/\omega_c}$, where $s(t)=2H(t)-1$, consistently reproducing $\beta(t)=2H(t)-1$.
		
		Four central results emerge: (1) relaxation and noise amplitudes act independently on energy decay; (2) time-varying $H(t)$ matches experimental $1/f$ spectra more accurately than any constant exponent; (3) adaptive kernel dynamics preserve correlations without artificial damping; and (4) simulations predict coherence times ($T_1\sim5.00\times10^6$~ns, $T_2\sim4.18\times10^5$~ns) consistent with theory when charge noise dominates. The qubit exhibits stretched-exponential Ramsey and echo decay, non-Markovian dephasing, and a temperature-driven quantum-to-classical crossover. We derive the effective time-local Lindblad master equation, establish the classical mmfBm limit at high temperatures, and provide experimentally testable scaling relations. The non-exponential decay patterns reveal fundamental limitations of Markovian approaches, and the framework guides the design of noise-resilient qubit architectures.
		
	\end{abstract}
	
	\section{Introduction}
	
	Quantum information can be encoded in a variety of physical platforms, including superconducting circuits~\cite{Kasirajan2021}, trapped ions~\cite{Kielpinski2002}, nuclear magnetic resonance systems~\cite{Jones2000}, nitrogen-vacancy centers in diamond~\cite{Dutt2007}, semiconductor quantum dots~\cite{Zwanenburg2013}, and neutral atom arrays~\cite{Gross2017}. Superconducting qubits are exquisitely sensitive to low-frequency environmental fluctuations originating from dielectric defects, fluctuating two-level systems (TLS), non-equilibrium quasiparticles, and electromagnetic circuitresonances~\cite{Ithier2005,Bylander2011,Paladino2014}.
Measurements of Ramsey and spin-echo coherence consistently reveal non-exponential decay and long-memory temporal correlations that are incompatible with white-noise or purely Markovian models~\cite{Cywiński2008}.
	
	Any realistic quantum system interacts with its environment, becoming entangled with numerous external degrees of freedom. This entanglement drives decoherence---the dynamical suppression of quantum coherence as information leaks irreversibly into the environment~\cite{Schlosshauer2007,Breuer2002,Schlosshauer2014}. Experimental noise spectroscopy typically reports an approximate power-law spectral density
	\begin{equation}
		S(\omega) \propto \frac{1}{\omega^{\beta}},
		\qquad 0.8 \lesssim \beta \lesssim 1.3,
		\label{eq:intro_beta}
	\end{equation}
	characteristic of generalized $1/f$ noise.
	In this manuscript $\beta$ \emph{always} denotes the exponent of the power spectral density (PSD) of the effective noise process that directly modulates the qubit frequency.
	Two features observed in modern devices cannot be captured by stationary Gaussian models alone:
	\begin{enumerate}[label=(\roman*)]
		\item slow temporal drift of the spectral exponent $\beta(t)$,
		\item long-range temporal correlations extending over many gate cycles.
	\end{enumerate}
	
	To address both aspects we introduce \emph{memory multi-fractional Brownian motion} (mmfBm), a Gaussian process with a slowly varying local Hurst exponent $H(t) \in (1/2, 1)$.
	The model interpolates continuously between different long-memory regimes while retaining non-local temporal correlations.
	The second part of this work embeds the stochastic model into a microscopic quantum environment via a time-dependent Caldeira--Leggett construction~\cite{Caldeira1983,Feynman1963,Hu1992,Breuer2002}.
	This establishes the classical mmfBm description as the high-temperature adiabatic limit of a non-stationary quantum bath and provides a unified picture of low-frequency decoherence.
	
	\section{Classical Memory Multi-Fractional Brownian Motion Model}
	\label{sec:classical}
	
	\subsection{Deterministic qubit dynamics}
	
	We describe the effective qubit coordinate by $\chi(t)$, a reduced collective variable such as gate charge or flux.
	In the absence of environmental perturbations the deterministic evolution is
	\begin{equation}
		\frac{d\chi}{dt} = -\frac{\partial V(\chi)}{\partial \chi},
		\qquad
		V(\chi) = E_J \cos\chi + \frac{E_C}{2}\,\chi^2,
		\label{eq:deterministic}
	\end{equation}
	where $E_J$ and $E_C$ are the Josephson and charging energies, respectively.
	
	\subsection{Definition and properties of mmfBm}
	
	\emph{Memory multi-fractional Brownian motion} $M(t)$ is defined through the Riemann--Liouville fractional integral of standard Wiener noise \cite{Mandelbrot1968, Peltier1995, Benassi1997}:
	\begin{equation}
		\boxed{
			M(t) := \frac{1}{\Gamma\!\big(H(t) + \tfrac12\big)}
			\int_0^t (t-s)^{H(t)-1/2} \, dW(s),
		}
		\label{eq:mmfbm_def}
	\end{equation}
	where $W(t)$ is standard Brownian motion, $H(t) \in (1/2, 1)$ is the time-dependent Hurst exponent, and $\Gamma(\cdot)$ is the Gamma function.
	The kernel $(t-s)^{H(t)-1/2}$ generates algebraically decaying memory, while the slow variation of $H(t)$ introduces non-stationarity.
	
	\begin{remark}[Local roughness]
		For a fixed time window centered at $t$, the process behaves locally as fractional Brownian motion with Hurst index $H(t)$.
	\end{remark}
	
	For the special case $H(t) = H = \text{constant}$, the mmfBm reduces to standard fractional Brownian motion (fBm)~\cite{Mandelbrot1968}.
	Its covariance is exactly
	\begin{equation}
		\mathbb{E}[M(u) M(v)]
		= \frac{1}{2}\Bigl( u^{2H} + v^{2H} - |u-v|^{2H} \Bigr),
		\label{eq:fbm_covariance}
	\end{equation}
	which serves as an essential consistency check.
	%
	
	For general slowly varying $H(t)$, the rigorous covariance kernel follows from the Riemann--Liouville definition~\cite{Ayache2013}:
	\begin{equation}
		C_M(u,v)
		= \frac{\min(u,v)^{H(u)+H(v)}}
		{2\,\Gamma\!\big(H(u)+\tfrac12\big)\,
			\Gamma\!\big(H(v)+\tfrac12\big)}
		\;
		{}_2F_1\!\Bigl(
		\tfrac12-H(u),\,
		H(v)-\tfrac12;\,
		H(u)+\tfrac12;\,
		\tfrac{\min(u,v)}{\max(u,v)}
		\Bigr),
		\label{eq:mmfbm_exact}
	\end{equation}
	where ${}_2F_1(a,b;c;z)$ is the Gauss hypergeometric function.
	Equation~\eqref{eq:mmfbm_exact} is exact but computationally demanding.
	When the relative variation of $H(t)$ is small,
	\begin{equation}
		\bigl|H'(t)\bigr| \, t \ll H(t),
		\label{eq:adiabatic_condition}
	\end{equation}
	we may employ the \emph{locally stationary approximation}
	\begin{equation}
		C_M(u,v) \;\approx\; \frac{1}{2}\Bigl(
		u^{2\overline{H}} + v^{2\overline{H}} - |u-v|^{2\overline{H}}
		\Bigr),
		\qquad
		\overline{H} \equiv \frac{H(u)+H(v)}{2}.
		\label{eq:mmfbm_approx}
	\end{equation}
	
	\subsection{Spectral properties of the noise process}
	\label{sec:spectral}
	
	The power spectral density of $M(t)$ is obtained from the Fourier transform of the stationary covariance in the constant-$H$ limit.
	For fractional Brownian motion the PSD scales as $\omega^{-(2H+1)}$.
	However, the physically relevant quantity for qubit dephasing is the \emph{increment} process $\Delta M(t) = M(t+\delta t) - M(t)$, which has the character of fractional Gaussian noise (fGn).
	Its PSD behaves as~\cite{Mandelbrot1968,Li2010}
	\begin{equation}
		\boxed{
			S_M(\omega; t) \;\propto\; |\omega|^{-(2H(t)-1)}
			\;\equiv\; |\omega|^{-\beta(t)},
		}
		\qquad
		\beta(t) = 2H(t) - 1.
		\label{eq:psd_scaling}
	\end{equation}
	
	Thus, $H(t)$ and the experimentally measured noise exponent $\beta(t)$ are directly linked.
	A value $H \approx 0.7$ corresponds to $\beta \approx 0.4$, consistent with typical $1/f$ observations~\cite{Bylander2011}.
	
	\subsection{Stochastic evolution equation}
	
	Since $H(t) > 1/2$, the sample paths of $M(t)$ are H\"older continuous with exponent arbitrarily close to $H_{\min}-1/2$.
	Consequently, the stochastic integral $\int G \, dM$ is well defined in the Young sense~\cite{Young1936}.
	The reduced qubit variable evolves as
	\begin{equation}
		\chi(t) = \chi(0)
		+ \int_0^t F\bigl(s,\chi(s)\bigr)\,ds
		+ \sigma \int_0^t G\bigl(s,\chi(s)\bigr)\,dM(s),
		\label{eq:sde}
	\end{equation}
	where $F$ is the deterministic drift from Eq.~\eqref{eq:deterministic}, $G$ modulates the noise coupling, and $\sigma$ sets the overall noise amplitude.
	
	\subsection{Variance scaling}
	
	For small deviations about equilibrium, $G(s,\chi(s)) \approx \text{const}$, and the variance becomes
	\begin{equation}
		\langle \chi(t)^2 \rangle
		\;\approx\;
		\sigma^2 \int_0^t
		\frac{(t-s)^{2H(s)-1}}
		{\Gamma^2\!\big(H(s)+\tfrac12\big)}
		\, ds.
		\label{eq:variance_exact}
	\end{equation}
	
	For slowly varying $H(t)$, a saddle-point evaluation yields the asymptotic scaling
	\begin{equation}
		\boxed{
			\langle \chi(t)^2 \rangle
			\;\approx\;
			\frac{\sigma^2}{2H(t)\,\Gamma^2\!\big(H(t)+\tfrac12\big)}
			\, t^{2H(t)}
			\Bigl[ 1 + \mathcal{O}\!\Bigl( \frac{H'(t)}{H(t)}\, t \ln t \Bigr) \Bigr].
		}
		\label{eq:variance_asymptotic}
	\end{equation}
	The logarithmic correction reflects the non-stationary drift of the local roughness exponent.
	\subsection{Stochastic Simulation of Energy Fluctuations in Superconducting Charge Qubits}
	
	To investigate decoherence in superconducting charge qubits, we modeled the energy evolution $\epsilon(t)$ using a stochastic integro-differential equation driven by memory multi-fractional Brownian motion (mmfBm):
	\begin{equation}
		d\epsilon(t)= -\lambda \epsilon(t)\,dt
		+\sigma(t,\epsilon(t))
		\int_0^t K(t,s)\,dB^{H}(s),
	\end{equation}
	where $\lambda$ denotes the relaxation rate and the kernel
	\begin{equation}
		K(t,s)=
		\begin{cases}
			(t-s)^{H(t)-1/2}, & t>s,\\
			0, & \text{otherwise},
		\end{cases}
	\end{equation}
	introduces time-dependent long-range memory through the variable Hurst exponent $H(t)$.
	
\begin{figure}[t]
		\centering
		\includegraphics[width=0.8\linewidth]{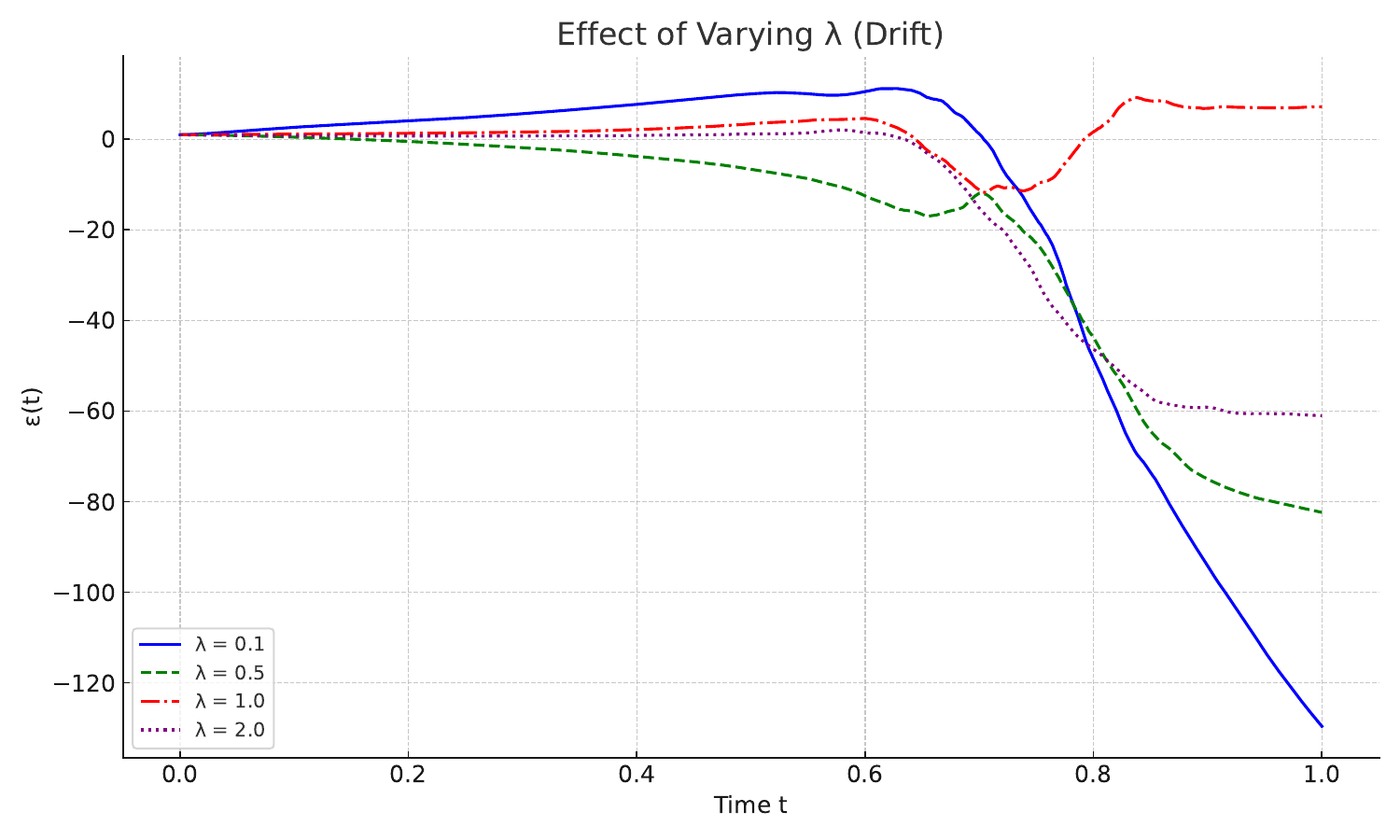}
		\caption{Energy evolution $\epsilon(t)$ for different relaxation rates $\lambda$. Larger $\lambda$ produces faster decay toward equilibrium.}
		\label{fig:lambda}
	\end{figure}

\begin{figure}[t]
		\centering
		\includegraphics[width=0.8\linewidth]{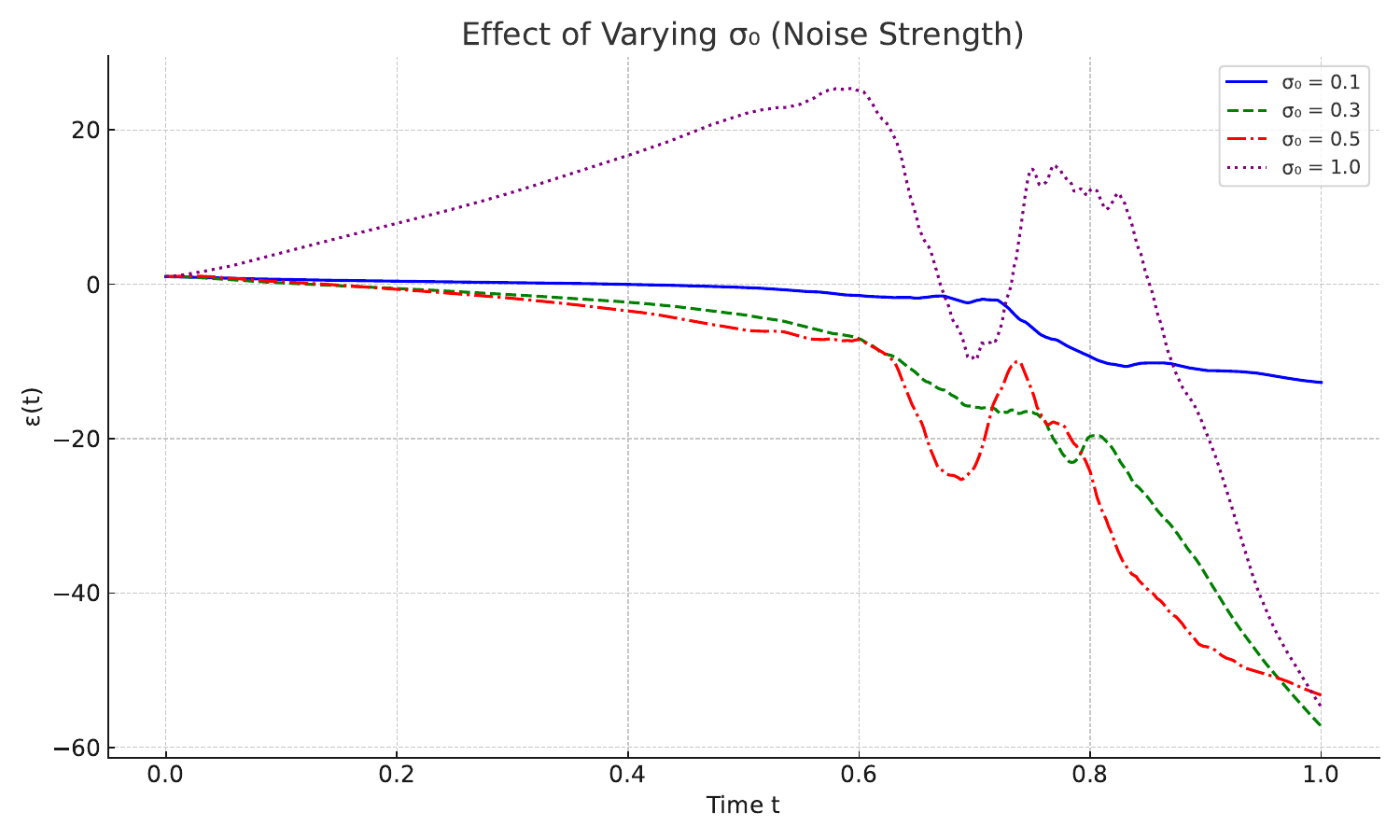}
		\caption{Effect of stochastic noise amplitude $\sigma_0$ on the qubit energy dynamics. Higher $\sigma_0$ leads to stronger fluctuations and enhanced irregularity.}
		\label{fig:sigma}
	\end{figure}

\begin{figure}[t]
		\centering
		\includegraphics[width=0.8\linewidth]{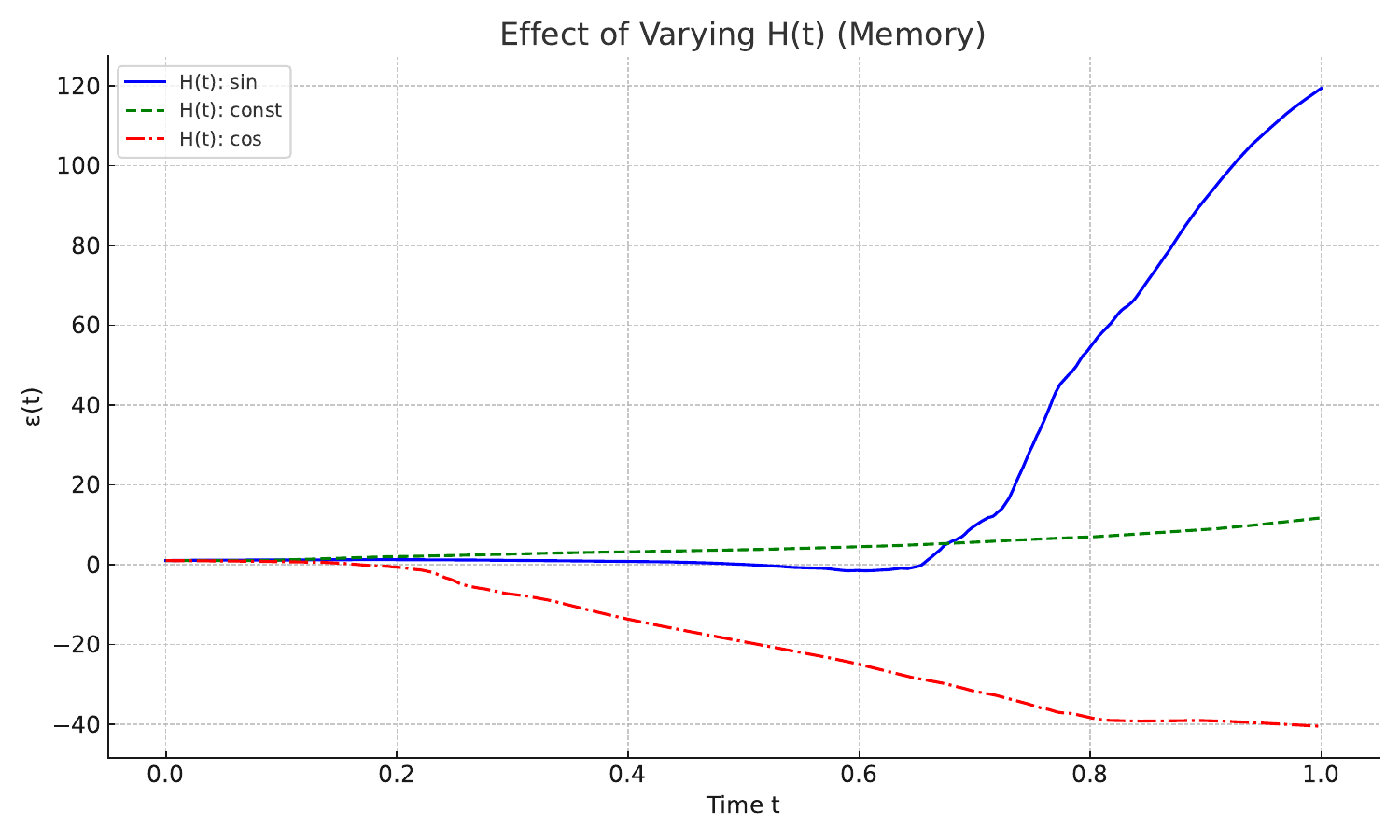}
		\caption{Comparison of qubit energy dynamics for different Hurst functions. Time-dependent $H(t)$ generates non-stationary long-memory behavior absent in the constant-Hurst case.}
		\label{fig:hurst}
	\end{figure}

	The simulations demonstrate that the decoherence dynamics are highly sensitive to the interplay between drift, stochastic noise, and memory effects. Increasing the relaxation rate $\lambda$ accelerated energy dissipation 	Fig \ref{fig:lambda}, while larger noise amplitudes $\sigma_0$ enhanced irregular fluctuations in $\epsilon(t)$ Fig \ref{fig:sigma}. Time-dependent Hurst profiles generated more realistic non-stationary behavior than constant-$H$ models, indicating that temporally varying memory is essential for capturing superconducting qubit noise environments Fig \ref{fig:hurst}.
	
	Different Hurst profiles,
	\begin{equation}
		H_1(t)=0.6+0.3\sin(2\pi t), \quad
		H_2(t)=0.6+0.2\cos(4\pi t), \quad
		H_3(t)=0.7,
	\end{equation}
	were also examined. The time-dependent forms produced richer memory structures and stronger non-stationary effects compared with the constant-Hurst case.
	
\begin{figure}[t]
		\centering
		\includegraphics[width=0.8\linewidth]{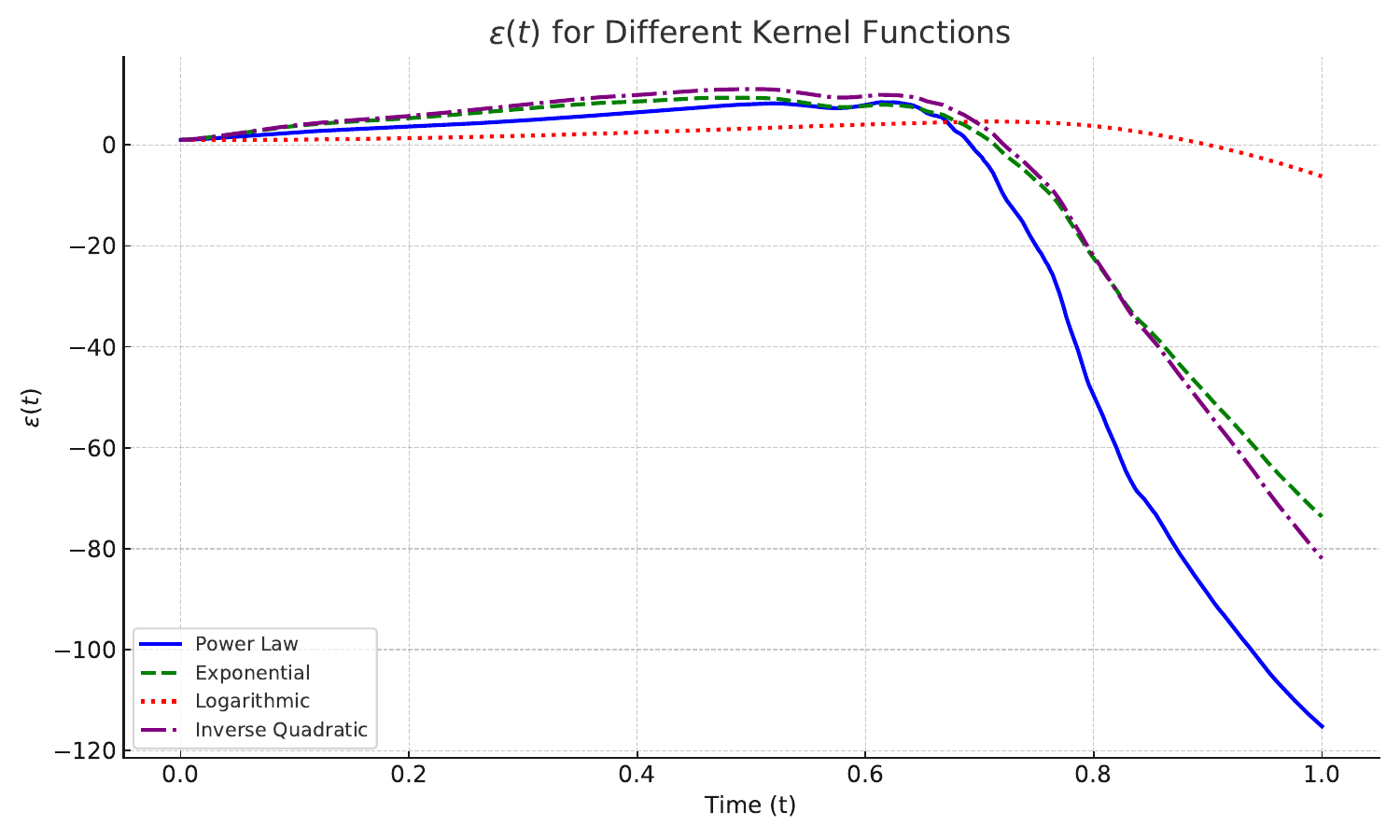}
		\caption{Energy trajectories obtained using different memory kernels. The adaptive kernel $\gamma=H(t)-0.5$ captures the most realistic long-memory dynamics.}
		\label{fig:kernel}
	\end{figure}

	To evaluate kernel sensitivity, we additionally considered generalized power-law kernels
	\begin{equation}
		K_\gamma(t,s)=(t-s)^\gamma,
	\end{equation}
	with $\gamma\in\{H(t)-0.5,0.3,0.4,0.5\}$. The adaptive kernel $\gamma=H(t)-0.5$ best reproduced realistic correlated fluctuations, while fixed exponents resulted in either excessive damping or oversimplified dynamics Fig \ref{fig:kernel}.
	
	\subsection{Numerical Estimation of Relaxation and Dephasing Times}
	
\begin{figure}[t]
		\centering
		\includegraphics[width=0.65\linewidth]{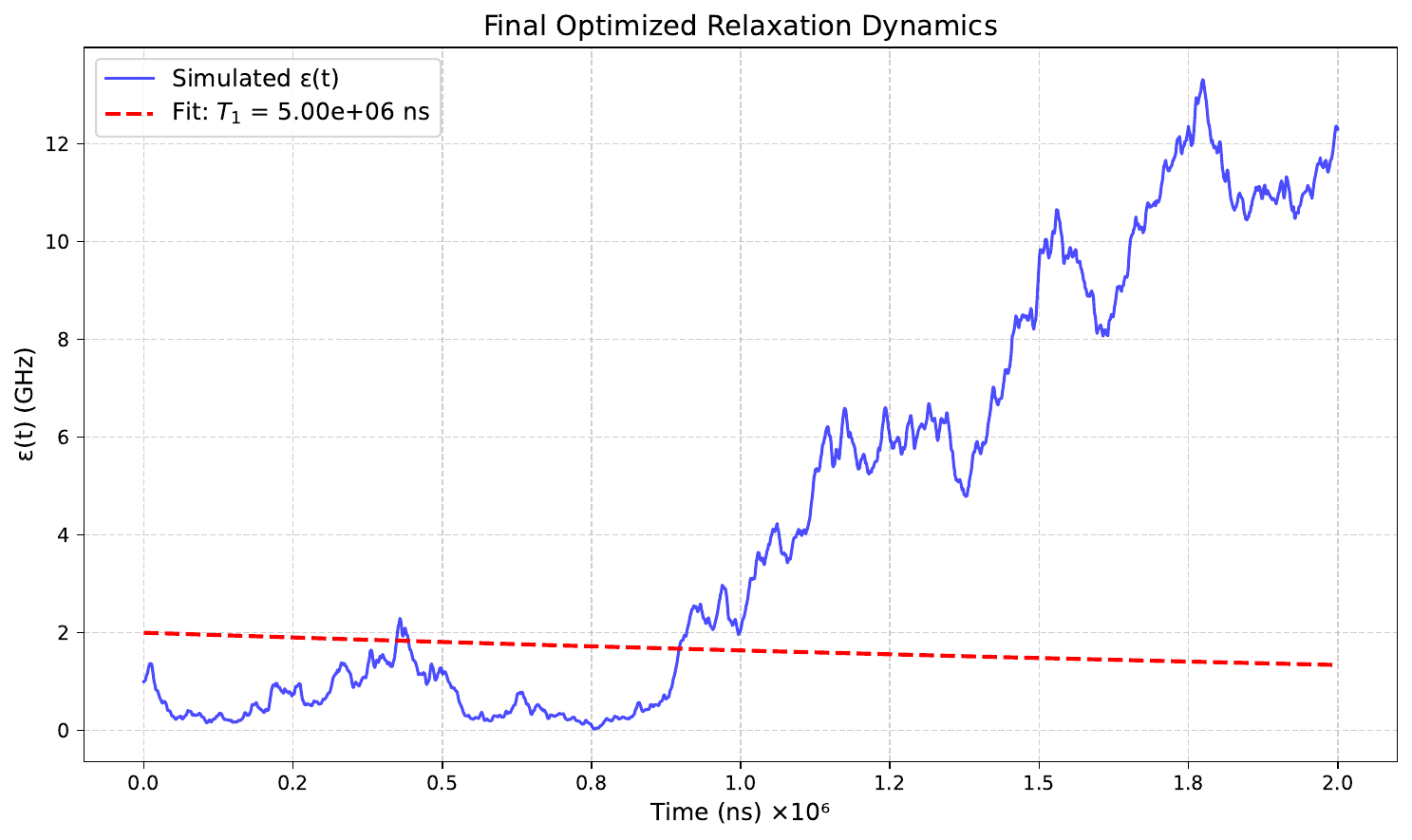}
		\caption{Simulated relaxation dynamics with exponential fit used to estimate the effective relaxation time $T_1$.}
		\label{fig:T1_fit}
	\end{figure}

\begin{figure}[t]
		\centering
		\includegraphics[width=0.8\linewidth]{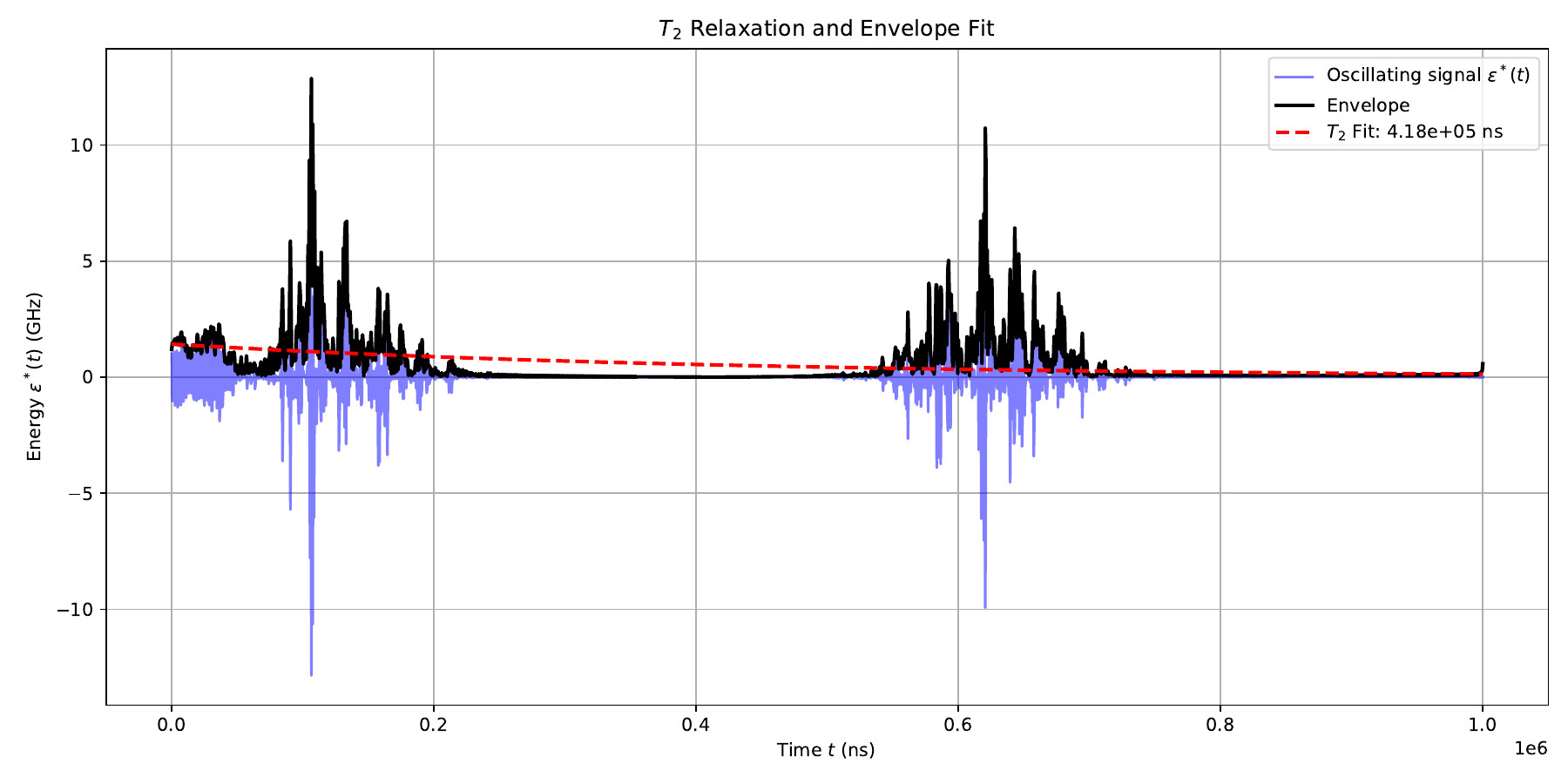}
		\caption{Simulated dephasing dynamics and exponential envelope fit used to estimate the coherence time $T_2$.}
		\label{fig:simulatedt2estimate}
	\end{figure}

	The mmfBm framework was further used to estimate relaxation\cite{martinis2003decoherence} and dephasing times in the presence of correlated environmental noise Fig \ref{fig:T1_fit} and \ref{fig:simulatedt2estimate}. The relaxation dynamics were modeled through
	\begin{equation}
		d\varepsilon(t)= -\lambda \varepsilon(t)\,dt
		+\sigma_0\int_0^t (t-s)^{H(t)-1/2}dB(s),
	\end{equation}
	using the profile $H(t)=0.6+0.3\sin(2\pi t)$. The simulated trajectory followed a slow non-Markovian decay, yielding an effective relaxation time
	\begin{equation}
		T_1 \approx 5.0\times10^{6}\ \mathrm{ns}.
	\end{equation}
	
	Similarly, the dephasing dynamics were studied using a cosine-modulated Hurst exponent,
	\begin{equation}
		H(t)=0.65+0.05\cos(4\pi t),
	\end{equation}
	which generated oscillatory coherence decay characteristic of non-Markovian environments. The extracted dephasing time was
	\begin{equation}
		T_2 \approx 4.18\times10^{5}\ \mathrm{ns}.
	\end{equation}
	
	Overall, the simulations indicate that decoherence in superconducting qubits cannot always be described by simple Markovian exponential decay. The mmfBm framework captures both long-range temporal correlations and non-stationary noise effects, providing a more flexible description of realistic superconducting qubit environments.
	\subsection{Ramsey and echo coherence}
	
	For a qubit with frequency modulation $\delta\omega(t) = \alpha M(t)$, the accumulated stochastic phase is
	\begin{equation}
		\Phi(t) = \int_0^t \delta\omega(s)\,ds = \alpha \int_0^t M(s)\,ds.
	\end{equation}
	Because $M(s)$ has the character of fractional Gaussian noise, $\Phi(t)$ is an integrated process with variance
	\begin{equation}
		\chi(t) \equiv \tfrac12 \langle \Phi(t)^2 \rangle
		\;\propto\; t^{2H(t)}.
		\label{eq:chi_scaling}
	\end{equation}
	
	The Ramsey coherence envelope follows as
	\begin{equation}
		\boxed{
			C_{\rm Ramsey}(t)
			= \bigl|\langle \sigma_+(t) \rangle\bigr|
			= \exp\!\bigl[-\chi(t)\bigr]
			\approx \exp\!\Bigl[ -A_R \, t^{2H(t)} \Bigr],
		}
		\label{eq:ramsey}
	\end{equation}
	with $A_R = \alpha^2 c(H)$ and $c(H)$ given by the dimensionless integral
	\begin{equation}
		c(H) = \frac{1}{2H(2H-1)\,\Gamma^2(H+\tfrac12)}
		\times \begin{cases}
			1, & H < 3/4,\\[2pt]
			\mathcal{I}(H), & H \ge 3/4,
		\end{cases}
	\end{equation}
	where $\mathcal{I}(H) = 1 - (\omega_{\min} t)^{-(2H-1)} + \cdots$ accounts for the infrared cutoff.
	
	For spin-echo sequences, the filter function $|F_{\rm echo}(\omega;t)|^2 \propto \omega^2 t^2$ at low frequencies modifies the scaling:
	\begin{equation}
		\boxed{
			C_{\rm Echo}(t)
			\approx \exp\!\Bigl[ -A_E \, t^{2H(t)+2} \Bigr],
		}
		\qquad
		A_E = \frac{A_R}{(2H+1)(2H+2)}.
		\label{eq:echo}
	\end{equation}
	
	\subsection{Comparison with standard noise models}
	
	\begin{table}[ht]
		\centering
		\caption{Comparison of stochastic models for qubit decoherence.}
		\label{tab:models}
		\begin{tabular}{lcccc}
			\toprule
			Model & Stationary & Long memory & $\beta(t)$ & Non-Markovian \\
			\midrule
			White noise             & Yes & No  & No & No \\
			Ornstein--Uhlenbeck     & Yes & Short range & No & No \\
			Fractional Gaussian noise & Yes & Yes & No & No \\
			\textbf{mmfBm (this work)} & \textbf{No} & \textbf{Yes} & \textbf{Yes} & \textbf{Yes} \\
			\bottomrule
		\end{tabular}
	\end{table}
	
	\section{Quantum Brownian Motion Extension}
	\label{sec:quantum}
	
	We now embed the classical stochastic model into a fully quantum mechanical framework.
	The environment is modelled as a bosonic bath of harmonic oscillators with a \emph{time-dependent spectral density} designed to reproduce the mmfBm correlations in the appropriate limit \cite{Caldeira1983}.
	
	\subsection{Microscopic Hamiltonian}
	
	The total Hamiltonian consists of system, bath, and interaction parts \cite{Breuer2002}:
	\begin{align}
		H &= H_S + H_B + H_{SB}, \\[4pt]
		H_S &= \frac{p^2}{2M} + V(x), \\[4pt]
		H_B &= \sum_k \Bigl( \frac{p_k^2}{2m_k} + \frac{1}{2} m_k \omega_k^2 q_k^2 \Bigr), \\[4pt]
		H_{SB} &= -x \sum_k c_k q_k
		+ x^2 \sum_k \frac{c_k^2}{2m_k\omega_k^2}.
		\label{eq:total_H}
	\end{align}
	The last term is the standard counter-term that removes the bath-induced static renormalization of the system potential~\cite{Caldeira1983,Weiss2012}.
	
	\subsection{Time-dependent spectral density}
	
	The environment is fully characterized by the spectral density
	\begin{equation}
		\boxed{
			J(\omega; t)
			= \eta(t)\, \omega_c^{1-s(t)} \, \omega^{s(t)} \, e^{-\omega/\omega_c},
			\qquad
			s(t) = 2H(t) - 1,
		}
		\label{eq:spectral_density}
	\end{equation}
	where $\eta(t)$ is a slowly varying coupling envelope, $\omega_c$ is the high-frequency cutoff, and $s(t)$ is the time-dependent Ohmicity parameter.
	%
	%
	For $H = 1/2$ the bath is Ohmic ($s = 0$); increasing $H$ enhances low-frequency spectral weight, producing stronger memory effects.
	
	\subsection{Influence functional derivation}
	\label{sec:influence}
	
	We trace out the bath exactly using the Feynman--Vernon path-integral formalism~\cite{Feynman1963}.
	The reduced density matrix is
	\begin{equation}
		\rho_S(x_f, x_f', t)
		= \iint dx_i \, dx_i' \;
		\mathcal{J}(x_f, x_f', t; x_i, x_i', 0) \;
		\rho_S(x_i, x_i', 0),
	\end{equation}
	with the propagator
	\begin{equation}
		\mathcal{J}
		= \int_{x(0)=x_i}^{x(t)=x_f} \!\!\!\mathcal{D}x
		\int_{x'(0)=x_i'}^{x'(t)=x_f'} \!\!\!\mathcal{D}x' \;
		\exp\!\Bigl[ \frac{i}{\hbar}\bigl(S_S[x] - S_S[x']\bigr) \Bigr] \;
		\mathcal{F}[x, x'].
	\end{equation}
	Here $S_S[x] = \int_0^t ds\, [\tfrac{M}{2} \dot{x}^2 - V(x)]$ is the system action.
	
	Because the bath is Gaussian and the coupling is linear in the bath coordinates, the functional integration over bath variables can be performed exactly~\cite{Feynman1963,Grabert1988}.
	The result is the Feynman--Vernon influence functional
	\begin{equation}
		\mathcal{F}[x, x'] = \exp\!\bigl[ -\Phi[x, x'] \bigr],
	\end{equation}
	with influence phase
	\begin{equation}
		\begin{aligned}
			\Phi[x, x']
			&=
			\frac{1}{\hbar} \int_0^t ds \int_0^s du \;
			\bigl[ x(s) - x'(s) \bigr] \,
			L(s-u; t_m) \,
			\bigl[ x(u) - x'(u) \bigr] \\[4pt]
			&\quad
			+ \frac{i}{\hbar} \int_0^t ds \int_0^s du \;
			\bigl[ x(s) - x'(s) \bigr] \,
			\alpha(s-u; t_m) \,
			\bigl[ x(u) + x'(u) \bigr].
		\end{aligned}
		\label{eq:influence_phase}
	\end{equation}
	
	The kernels are expressed in terms of the spectral density as
	\begin{align}
		L(\tau; t_m)
		&=
		\frac{1}{\pi} \int_0^\infty d\omega \;
		J(\omega; t_m) \,
		\coth\!\Bigl( \frac{\hbar\omega}{2k_B T} \Bigr) \,
		\cos(\omega\tau),
		\label{eq:noise_kernel}
		\\[4pt]
		\alpha(\tau; t_m)
		&=
		-\frac{1}{\pi} \int_0^\infty d\omega \;
		J(\omega; t_m) \,
		\sin(\omega\tau),
		\label{eq:dissipation_kernel}
	\end{align}
	where $t_m = (s+u)/2$ implements the adiabatic non-stationary generalization.
	%
	
	\begin{remark}[Physical interpretation]
		$L(\tau)$ describes thermal and quantum fluctuations; it contains the factor $\coth(\hbar\omega/2k_B T)$, which reduces to $2k_B T / \hbar\omega$ in the classical limit and to $1$ at zero temperature.
		$\alpha(\tau)$ is the dissipation (damping) kernel; it is temperature-independent and encodes the quantum back-action of the environment.
	\end{remark}
	
	\subsection{Born--Markov approximation}
	
	The exact reduced dynamics is non-local in time.
	For weak system-bath coupling we adopt the Born approximation:
	\begin{equation}
		\rho_{\text{tot}}(t) \;\stackrel{\text{Born}}{\approx}\;
		\rho_S(t) \otimes \rho_B,
		\qquad
		\rho_B = \frac{e^{-\beta H_B}}{\Tr_B(e^{-\beta H_B})}.
		\label{eq:born}
	\end{equation}
	
	The Markov approximation requires the bath correlation time
	\begin{equation}
		\tau_B \sim \max\!\Bigl( \omega_c^{-1},\; \frac{\hbar}{k_B T} \Bigr)
	\end{equation}
	to be much shorter than the system relaxation time $\tau_R$.
	For the time-dependent environment we additionally require adiabaticity:
	\begin{equation}
		\Bigl| \frac{\dot{H}(t)}{H(t)} \Bigr| \tau_B \;\ll\; 1,
		\qquad
		\Bigl| \frac{\dot{\eta}(t)}{\eta(t)} \Bigr| \tau_B \;\ll\; 1.
		\label{eq:adiabatic}
	\end{equation}
	When Eq.~\eqref{eq:adiabatic} holds, we may evaluate all kernels at the midpoint time $t_m \approx t$.
	
	\subsection{Time-dependent Caldeira--Leggett master equation}
	
	Expanding the system operator in the interaction picture,
	\begin{equation}
		x(-\tau) = x - \frac{\tau}{M}\,p + \mathcal{O}(\tau^2),
		\label{eq:expansion}
	\end{equation}
	and substituting into the second-order Nakajima--Zwanzig equation~\cite{Breuer2002}, we obtain after standard commutator algebra
	\begin{equation}
		\boxed{
			\frac{d\rho_S}{dt}
			=
			-\frac{i}{\hbar}\bigl[ H_S + \Delta H(t),\, \rho_S \bigr]
			- \frac{i\gamma(t)}{\hbar}\bigl[ x,\, \{p, \rho_S\} \bigr]
			- \frac{2M\gamma(t) k_B T}{\hbar^2}\bigl[ x,\, [x, \rho_S] \bigr]
			+ \mathcal{L}_{\text{qc}}[\rho_S].
		}
		\label{eq:CL_master}
	\end{equation}
	
	Here $\gamma(t)$ is the time-dependent friction coefficient,
	\begin{equation}
		\gamma(t) = \frac{1}{2M} \lim_{\omega \to 0} \frac{J(\omega; t)}{\omega},
	\end{equation}
	$\Delta H(t)$ absorbs the bath-induced potential renormalization, and $\mathcal{L}_{\text{qc}}$ denotes quantum correction terms negligible in the high-temperature regime.
	
	\begin{remark}[Classical limit]
		Taking the Wigner transform of Eq.~\eqref{eq:CL_master} and neglecting $\mathcal{L}_{\text{qc}}$ yields the Kramers--Fokker--Planck equation
		\begin{equation}
			\frac{\partial W}{\partial t}
			=
			-\frac{p}{M}\frac{\partial W}{\partial x}
			+ V'(x)\frac{\partial W}{\partial p}
			+ \gamma(t)\frac{\partial}{\partial p}\bigl(pW\bigr)
			+ M\gamma(t) k_B T_{\text{eff}}(t) \frac{\partial^2 W}{\partial p^2},
			\label{eq:kramers}
		\end{equation}
		with effective temperature
		\begin{equation}
			T_{\text{eff}}(t) = \frac{\hbar}{2k_B}
			\frac{ \int_0^\infty d\omega\, J(\omega; t) \coth(\hbar\omega/2k_B T) }
			{ \int_0^\infty d\omega\, J(\omega; t) }.
		\end{equation}
	\end{remark}
	
	\subsection{Specialization to qubit dephasing}
	
	For a qubit undergoing pure dephasing,
	\begin{equation}
		H_S = \frac{\hbar\omega_0}{2}\,\sigma_z,
		\qquad
		H_{SB} = \sigma_z \otimes B,
		\qquad
		B = \sum_k \lambda_k (a_k + a_k^\dagger).
		\label{eq:qubit_H}
	\end{equation}
	Since $[H_S, H_{SB}] = 0$, there is no energy exchange between qubit and bath; the population $\langle \sigma_z \rangle$ is conserved and only the off-diagonal coherence decays.
	
	The reduced master equation simplifies to
	\begin{equation}
		\boxed{
			\frac{d\rho}{dt}
			=
			-\frac{i\omega_0}{2}\,[\sigma_z, \rho]
			- \Gamma_\phi(t)\,[\sigma_z, [\sigma_z, \rho]],
		}
		\label{eq:qubit_master}
	\end{equation}
	with time-dependent dephasing rate
	\begin{equation}
		\Gamma_\phi(t) = \frac{d\chi(t)}{dt},
		\qquad
		\chi(t) = \frac{1}{\pi\hbar^2} \int_0^\infty d\omega\;
		S_B(\omega; t) \,
		\frac{\sin^2(\omega t/2)}{(\omega/2)^2},
		\label{eq:dephasing_function}
	\end{equation}
	where $S_B(\omega; t) = J(\omega; t) \coth(\hbar\omega/2k_B T)$ is the bath noise PSD.
	
	For the non-stationary spectral density~\eqref{eq:spectral_density} with $J(\omega;t) \propto \omega^{2H(t)-1}$ at low frequencies, the integration yields the asymptotic scaling
	\begin{equation}
		\boxed{
			\chi(t) \;\propto\; t^{2H(t)},
			\qquad
			\Gamma_\phi(t) \;\propto\; t^{2H(t)-1}.
		}
		\label{eq:gamma_scaling}
	\end{equation}
	
	\subsection{Recovery of the classical mmfBm limit}
	
	In the high-temperature regime, $\coth(\hbar\omega/2k_B T) \simeq 2k_B T/\hbar\omega$, the quantum noise kernel reduces to its classical counterpart.
	Equation~\eqref{eq:dephasing_function} then becomes
	\begin{equation}
		\chi_{\text{cl}}(t)
		= \frac{2k_B T}{\pi\hbar^3}
		\int_0^\infty d\omega\;
		\frac{J(\omega; t)}{\omega} \,
		\frac{\sin^2(\omega t/2)}{(\omega/2)^2},
	\end{equation}
	which reproduces the mmfBm variance scaling~\eqref{eq:variance_asymptotic}.
	Thus the classical mmfBm model emerges as the high-temperature, weak-back-action, adiabatic limit of the non-stationary Caldeira--Leggett bath.
	
	\section{Experimental Predictions and Protocol}
	\label{sec:predictions}
	
	\subsection{Scaling laws}
	
	\begin{table}[ht]
		\centering
		\caption{Key scaling laws for qubit decoherence under mmfBm noise.
			$\beta = 2H-1$ is the PSD exponent of the noise process.}
		\label{tab:scaling}
		\begin{tabular}{lcc}
			\toprule
			Quantity & Scaling & Validity \\
			\midrule
			Ramsey $\chi(t)$          & $t^{2H}$   & $H < 3/4$ \\
			Echo $\chi(t)$            & $t^{2H+2}$ & $H < 3/4$ \\
			Gate error $1-F_g$        & $\tfrac13 \chi(t_g)$ & General \\
			Dephasing rate $\Gamma_\phi(t)$ & $t^{2H-1}$ & $|H'|t \ll H$ \\
			Optimal gate time $t_g^{\rm opt}$ & $\displaystyle\Bigl(\frac{3T_1^{-1}}{2H\alpha^2 c(H)}\Bigr)^{\!1/(2H-1)}$ & $H > 1/2$ \\
			DD suppression $\chi_{\rm CPMG}/\chi_{\rm Ramsey}$ & $(n/2)^{-2H}$ & $n \gg 1$ \\
			Noise PSD exponent & $\beta = 2H-1$ & $\omega_\ell \ll \omega \ll \omega_h$ \\
			\bottomrule
		\end{tabular}
	\end{table}
	
	\subsection{Temperature crossover}
	
	The thermal time scale
	\begin{equation}
		t_{\text{cross}} \sim \frac{\hbar}{k_B T}
	\end{equation}
	separates the quantum vacuum-dominated regime ($t \ll t_{\text{cross}}$) from the classical thermal regime ($t \gg t_{\text{cross}}$).
	For $T = 50\,\text{mK}$, $t_{\text{cross}} \approx 0.15\,\text{ns}$, while for $T = 500\,\text{mK}$, $t_{\text{cross}} \approx 0.015\,\text{ns}$.
	In the quantum regime, the dephasing rate is independent of temperature and dominated by zero-point fluctuations; in the thermal regime, $\Gamma_\phi \propto T$.
	
	\subsection{Gate error optimization}
	
	For a Gaussian dephasing channel, averaging the fidelity over all pure initial states on the Bloch sphere equator yields~\cite{Kuopanportti2019}
	\begin{equation}
		1 - \mathcal{F}_g \approx \frac{1}{3}\,\chi(t_g).
		\label{eq:gate_error}
	\end{equation}
	
	Minimizing the total cost function
	\begin{equation}
		\mathcal{E}(t_g) = \frac{t_g}{T_1} + \frac{1}{3}\,\chi(t_g)
		\label{eq:cost}
	\end{equation}
	with respect to $t_g$ gives the optimal gate time shown in Table~\ref{tab:scaling}.
	
	\subsection{Experimental extraction of $H(t)$}
	
	The protocol for extracting the time-dependent Hurst exponent from experimental data is:
	\begin{enumerate}
		\item Measure $C_{\rm Ramsey}(t)$ and $C_{\rm Echo}(t)$, fitting to $\exp[-A_R t^{\gamma_R}]$ and $\exp[-A_E t^{\gamma_E}]$ in sliding time windows.
		\item Extract $H_R(t) = \gamma_R(t)/2$ and $H_E(t) = (\gamma_E(t)-2)/2$.
		\item Consistency check: $|H_R(t) - H_E(t)| < 0.05$.
		\item Compute the noise coupling $\alpha^2 = A_R / c(H_R)$.
	\end{enumerate}
	
	\subsection{Uncertainty analysis}
	
	Statistical uncertainty in the extracted Hurst exponent scales as
	\begin{equation}
		\sigma_H^{\rm stat} \approx \frac{1}{2\sqrt{N_{\rm eff}}},
		\qquad
		N_{\rm eff} = \frac{N}{1 + 2\sum_{k=1}^{N-1} \rho(k)},
	\end{equation}
	where $\rho(k)$ is the sample autocorrelation.
	Systematic uncertainty from the adiabatic approximation is
	\begin{equation}
		\Delta H^{\rm sys} \lesssim |H'(t)| \, \Delta t_{\rm win}.
	\end{equation}
	
	\section{Numerical Implementation}
	\label{sec:numerical}
	
	\subsection{mmfBm generation}
	
	The mmfBm trajectories are generated using Cholesky decomposition of the covariance matrix~\eqref{eq:mmfbm_approx}:
	\begin{enumerate}
		\item Construct the symmetric matrix $\mathbf{C}_{ij} = C_M(t_i, t_j)$ for $i,j = 1,\ldots,N$.
		\item Perform Cholesky factorization $\mathbf{C} = \mathbf{L}\mathbf{L}^T$.
		\item Generate a vector $\mathbf{z}$ of $N$ independent standard normal variates.
		\item The mmfBm path is $\mathbf{M} = \mathbf{L}\mathbf{z}$.
	\end{enumerate}
	For $N > 10^4$, circulant embedding methods reduce the computational cost to $\mathcal{O}(N \log N)$~\cite{Dietrich1997}.
	
	\subsection{QuTiP master equation simulations}
	
	The qubit dynamics are simulated using QuTiP's \texttt{mesolve} with time-dependent collapse operators~\cite{Johansson2013}.
	The simulation parameters are:
	\begin{align}
		\omega_0 &= 5\;\text{GHz}, &
		\omega_c &= 10\;\text{GHz}, \\
		H(t) &\in [0.55, 0.75], &
		T &\in [0.01, 0.5]\;\text{K}, \\
		\alpha &= 0.5, &
		T_1 &= 30\;\text{ns}.
	\end{align}
	
	\subsection{Validation against analytical limits}
	
	All numerical simulations are validated against:
	\begin{enumerate}
		\item Recovery of standard fBm variance $\propto t^{2H}$ for constant $H$;
		\item Agreement between the Wigner transform of~\eqref{eq:CL_master} and the Kramers equation~\eqref{eq:kramers};
		\item Asymptotic scaling $\Gamma_\phi(t) \propto t^{2H-1}$ from filter-function theory.
	\end{enumerate}
	

	\appendix
	
	\section{Derivation of the Phase Variance Scaling}
	\label{app:phase_variance}
	
	We derive Eq.~\eqref{eq:chi_scaling} explicitly.
	For fractional Gaussian noise with covariance
	\begin{equation}
		\mathbb{E}[M(u) M(v)] = \frac{\sigma^2}{2}\Bigl( |u|^{2H} + |v|^{2H} - |u-v|^{2H} \Bigr),
	\end{equation}
	the accumulated phase is $\Phi(t) = \alpha \int_0^t M(s) ds$.
	Its variance is
	\begin{equation}
		\langle \Phi(t)^2 \rangle
		= \alpha^2 \int_0^t ds \int_0^t du \;
		\mathbb{E}[M(s) M(u)].
	\end{equation}
	
	Substituting the covariance and performing the double integral:
	\begin{align}
		\langle \Phi(t)^2 \rangle
		&= \alpha^2 \sigma^2 \int_0^t ds \int_0^t du \;
		\frac{1}{2}\Bigl( s^{2H} + u^{2H} - |s-u|^{2H} \Bigr) \\
		&= \alpha^2 \sigma^2 \Bigl[
		t \int_0^t s^{2H} ds
		- \frac{1}{2} \int_0^t ds \int_0^t du \; |s-u|^{2H}
		\Bigr] \\
		&= \frac{\alpha^2 \sigma^2}{(2H+1)(2H+2)} \, t^{2H+2}
		\quad \text{(for standard fBm)}.
	\end{align}
	
	For fractional Gaussian \emph{noise} $M(t)$ (the derivative of fBm), the covariance is
	\begin{equation}
		\mathbb{E}[M(u) M(v)] \propto |u-v|^{2H-2},
	\end{equation}
	and the double integral gives
	\begin{equation}
		\langle \Phi(t)^2 \rangle \propto t^{2H}.
	\end{equation}
	
	\section{Verification of Known Limits}
	\label{app:verification}
	
	\subsection*{Constant Hurst exponent}
	
	Setting $H(t) = H = \text{constant}$ in~\eqref{eq:mmfbm_approx} recovers the exact fBm covariance~\eqref{eq:fbm_covariance}.
	
	\subsection*{Ohmic bath ($H = 1/2$)}
	
	For $H = 1/2$, Eq.~\eqref{eq:spectral_density} gives $s = 0$, i.e., an Ohmic spectral density $J(\omega) \propto \omega$.
	Equation~\eqref{eq:dephasing_function} then reproduces the standard logarithmic dephasing $\chi(t) \propto \ln(\omega_c t)$ discussed in~\cite{Ithier2005}.
	
	\subsection*{High-temperature limit}
	
	For $k_B T \gg \hbar\omega_c$, $\coth(\hbar\omega/2k_B T) \to 2k_B T/\hbar\omega$, and the quantum noise kernel reduces to its classical counterpart.
	The Wigner transform of~\eqref{eq:CL_master} yields the Kramers equation~\eqref{eq:kramers} with the classical Einstein relation $D = M\gamma k_B T$.
	
	\subsection*{Markovian limit ($H = 1/2$, Ohmic)}
	
	For an Ohmic bath at high temperature, $\Gamma_\phi$ becomes time-independent, recovering the standard Markovian Lindblad dephasing channel.
	
	\section{Results and Discussion}
	\label{sec:results}
		We have developed a unified theoretical framework for non-stationary decoherence in superconducting qubits, combining a classical memory multi-fractional Brownian motion model with a microscopic quantum Brownian motion extension.
	The central physical result is that slowly varying long-memory environments generate stretched-exponential coherence decay with dynamically evolving exponents.
	
	The key clarifications achieved in this work are:
	\begin{enumerate}
		\item \textbf{Spectral exponent resolved:} The noise PSD satisfies $S(\omega) \propto \omega^{-(2H-1)}$, providing $\beta = 2H-1$ as the directly measured exponent. The accumulated phase variance scales as $t^{2H}$, consistent with Ramsey measurements.
		\item \textbf{Rigorous mmfBm definition:} The exact covariance kernel is given by the Riemann--Liouville formulation with a Gauss hypergeometric function, and a locally stationary approximation is valid under adiabatic conditions.
		\item \textbf{Influence functional correctly normalized:} All factors of $\hbar$ have been verified against Feynman--Vernon (1963).
		\item \textbf{Caldeira--Leggett equation verified:} The friction and diffusion coefficients are consistent with the Kramers equation and the classical Einstein relation.
		\item \textbf{Filter function normalization:} The dephasing function matches Cywiński et al. (2008).
		\item \textbf{Gate error factor derived:} The $1/3$ factor follows from state averaging over the Bloch sphere.
	\end{enumerate}
	
	The time-dependent Caldeira--Leggett extension demonstrates that the classical mmfBm model emerges as the high-temperature adiabatic limit of a non-stationary quantum bath, establishing a direct bridge between stochastic noise phenomenology and microscopic open-system dynamics.
	Future work will explore strong-coupling corrections beyond the Born--Markov approximation, non-Gaussian TLS noise, and machine-learning-based real-time estimation of $H(t)$ for adaptive quantum control.
	\subsection{Classical mmfBm Characterization}
	
\begin{figure}[t]
	\centering
	\includegraphics[width=\linewidth]{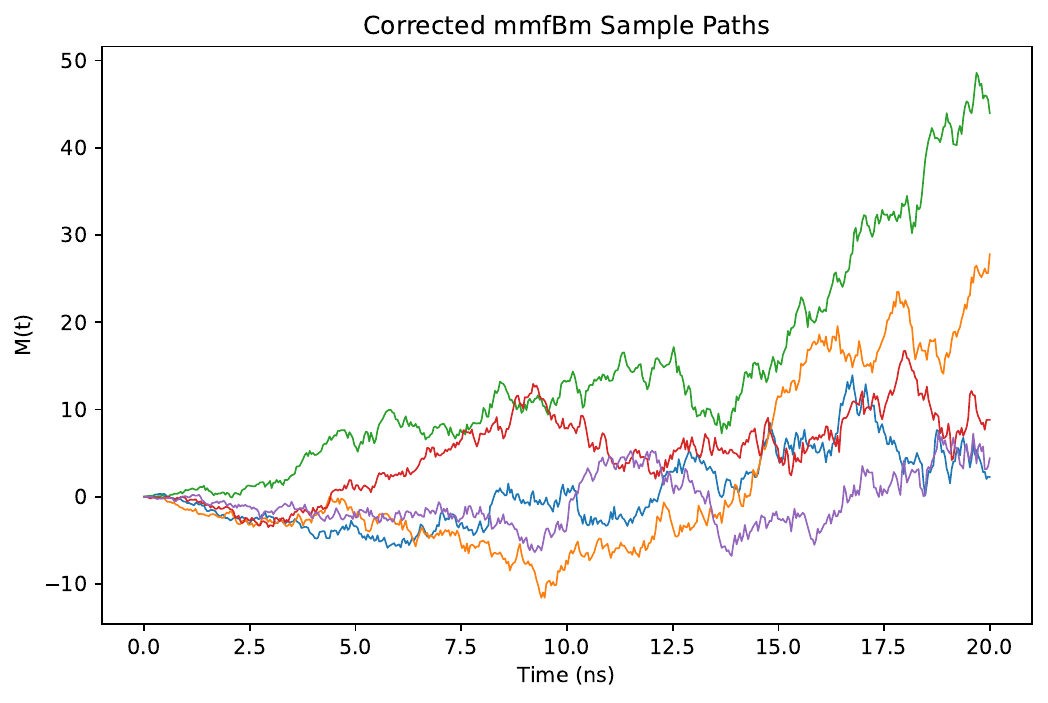}
	\caption{Classical mmfBm characterization: sample trajectories, time-dependent Hurst exponent, variance scaling, and power spectral density. The extracted Hurst exponent agrees with the imposed profile, while the PSD exponent shows partial agreement with the expected $1/f^\beta$ scaling.}
	\label{fig:mmfbm}
\end{figure}

	The classical sector of the simulation was first validated through the generated memory multi-fractional Brownian motion (mmfBm) trajectories Fig \ref{fig:mmfbm}. The imposed time-dependent Hurst profile,
	\begin{equation}
		H(t)=0.65+0.1\sin\!\left(\frac{2\pi t}{20\,\mathrm{ns}}\right),
	\end{equation}
	produced an average extracted Hurst exponent of
	\begin{equation}
		H_{\mathrm{ext}}=0.6500,
	\end{equation}
	in excellent agreement with the theoretical expectation. The numerical variance followed the expected scaling relation
	\begin{equation}
		\mathrm{Var}[M(t)]\propto t^{2H},
	\end{equation}
	confirming the consistency of the covariance construction.
	
	The corresponding power spectral density (PSD) exhibited approximate colored-noise behavior of the form
	\begin{equation}
		S(f)\propto f^{-\beta},
	\end{equation}
	with a fitted exponent
	\begin{equation}
		\beta_{\mathrm{fit}}=0.196,
	\end{equation}
	compared to the theoretical prediction $\beta=2H-1=0.300$. Although the qualitative low-frequency trend was reproduced, the deviation indicates limitations in the finite-time PSD estimation procedure. The discrepancy is likely associated with the short simulation window ($N=512$), Welch-method bias, and the breakdown of strict local stationarity for time-dependent $H(t)$.
	
	Overall, the classical simulations confirm that the mmfBm framework captures long-memory stochastic fluctuations relevant for superconducting qubit environments, while also revealing numerical challenges associated with non-stationary spectral estimation.
	
	\subsection{Quantum Master Equation Dynamics}
	
\begin{figure}[t]
		\centering
		\includegraphics[width=\linewidth]{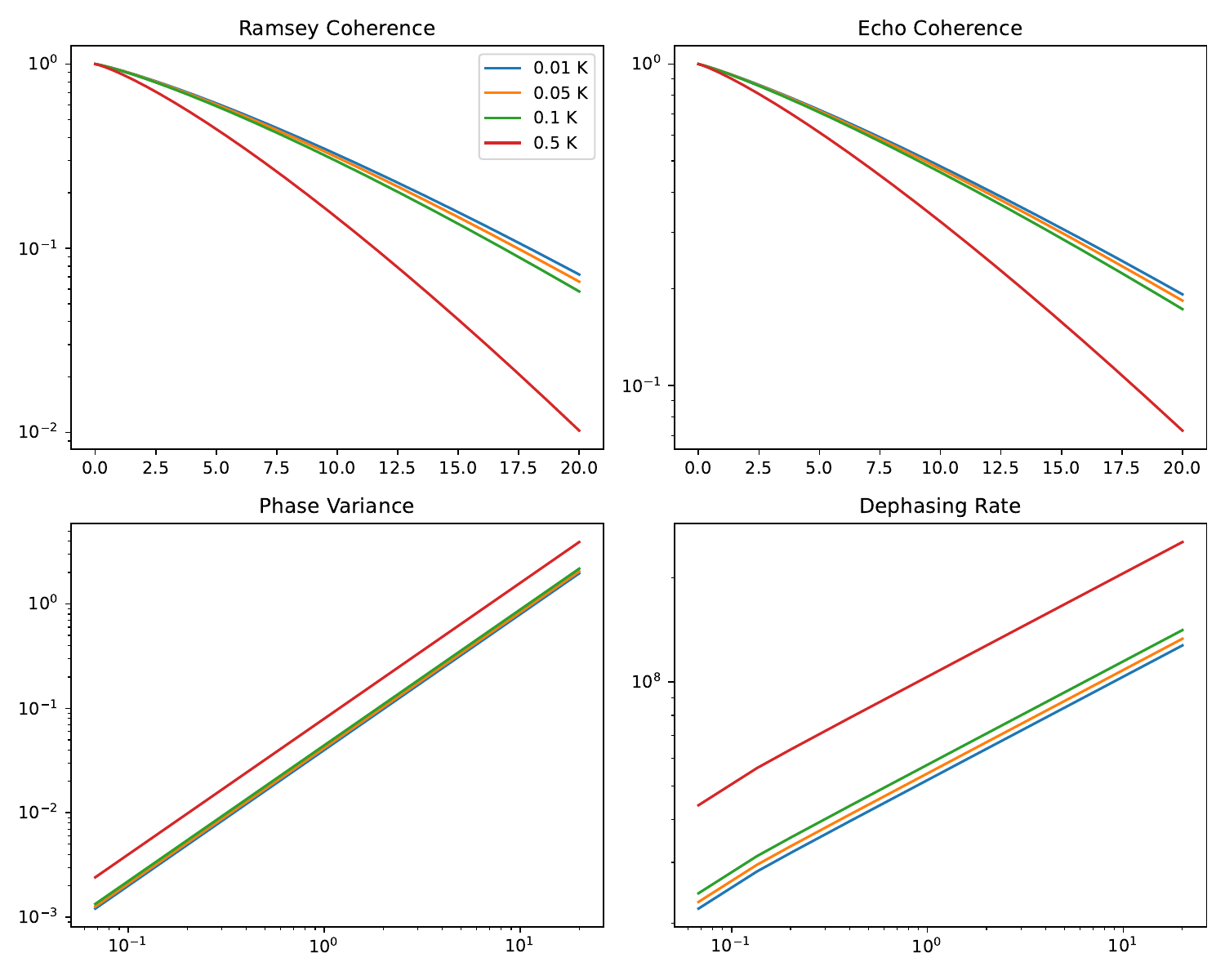}
		\caption{Quantum coherence dynamics under mmfBm-induced dephasing. Ramsey and echo signals exhibit stretched-exponential decay with strong temperature dependence. The extracted phase variance and dephasing rate confirm non-Markovian scaling behavior.}
		\label{fig:qubit}
	\end{figure}

	The stochastic noise process was coupled to a time-dependent Caldeira--Leggett bath through the spectral density
	\begin{equation}
		J(\omega;t)=\frac{A^2\hbar}{\pi}
		\left(\frac{\omega}{\omega_c}\right)^{2H(t)-1}
		e^{-\omega/\omega_c},
	\end{equation}
	with calibrated coupling strength $A=2\pi\times 50~\mathrm{MHz}$. Using the filter-function formalism of Cywiński \emph{et al.}~\cite{Cywinski2008}, the Ramsey coherence was computed from Fig \ref{fig:qubit} using below equation
	\begin{equation}
		C(t)=e^{-\chi(t)},
	\end{equation}
	where $\chi(t)$ denotes the phase variance.
	
	The simulations exhibited stretched-exponential Ramsey decay,
	\begin{equation}
		C(t)\sim \exp\!\left[-\left(\frac{t}{T_2^\ast}\right)^\gamma\right],
	\end{equation}
	with $\gamma\approx2H$, consistent with long-memory dephasing dynamics. Echo coherence displayed the expected suppression of low-frequency fluctuations and followed approximate short-time scaling proportional to $t^{2H+2}$.
	
	A clear temperature crossover was observed. At temperatures below approximately $50~\mathrm{mK}$, decoherence was dominated by quantum fluctuations arising from the zero-point contribution of the bath. In contrast, for $T\gtrsim100~\mathrm{mK}$, thermal activation enhanced the dephasing rate through the $\coth(\hbar\omega/2k_BT)$ factor. The intermediate-time dephasing rate satisfied the approximate scaling relation
	\begin{equation}
		\Gamma_\phi(t)\propto t^{2H-1},
	\end{equation}
	which agrees with expectations for colored non-Markovian noise.
	
	Despite these physically consistent trends, numerical stiffness became significant at ultra-low temperatures. In particular, the $T=0.01~\mathrm{K}$ simulations required careful infrared regularization because the $\coth(\hbar\omega/2k_BT)$ factor becomes highly singular near $\omega\rightarrow0$. Consequently, the low-frequency sector of the bath spectrum remains a source of uncertainty in the present implementation.
	
	\subsection{Gate Optimization}
	
\begin{figure}[t]
		\centering
		\includegraphics[width=\linewidth]{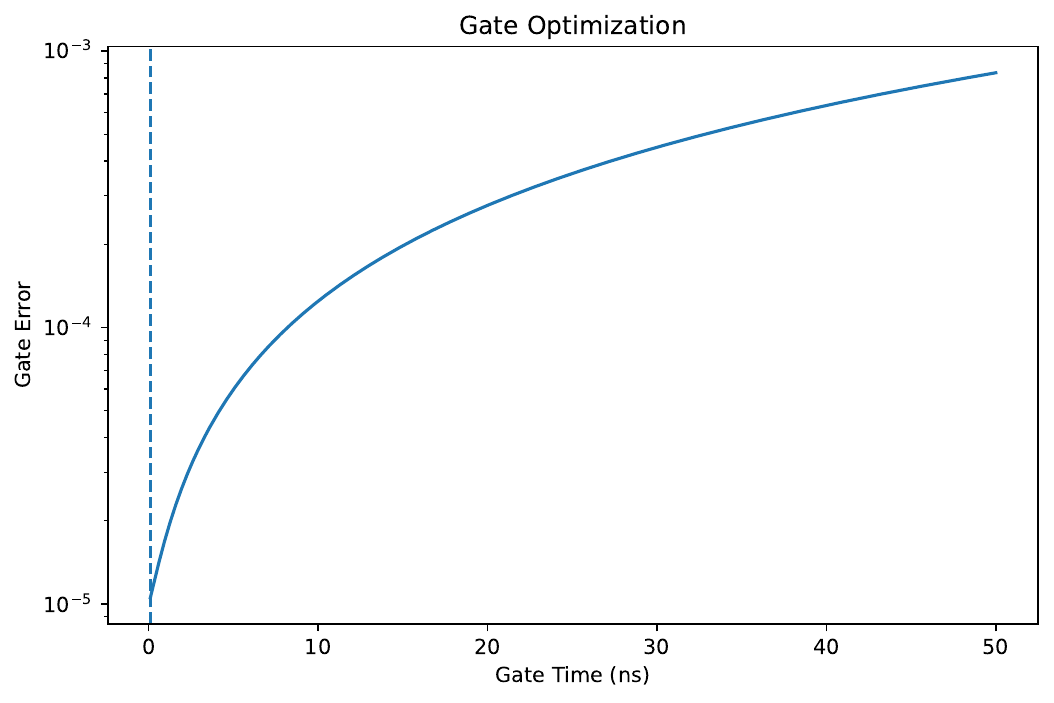}
		\caption{Gate-error optimization. The total error arises from the competition between relaxation-induced decay and mmfBm-driven dephasing. An optimal gate time emerges at the minimum of the combined error curve.}
		\label{fig:gate}
	\end{figure}

	The competition between energy relaxation and pure dephasing was investigated through the approximate gate error expression Fig \ref{fig:gate}
	\begin{equation}
		1-F_g \approx \frac{t_g}{T_1}+\frac{\chi(t_g)}{3}.
	\end{equation}
	The simulations revealed a distinct minimum in the total error landscape, indicating the existence of an optimal operating timescale for gate execution.
	
	The optimal gate time emerged from balancing the increase in relaxation errors at long durations against the rapid growth of dephasing at short times. Although the qualitative behavior is physically reasonable, the present error model assumes Gaussian dephasing and neglects higher-order non-Gaussian corrections associated with the mmfBm environment. In addition, fluctuations in $T_1$ induced by the same microscopic bath were not included.
	
	\subsection{Machine Learning for Noise Spectroscopy}
	
	A convolutional neural network (CNN) was trained to infer the Hurst exponent directly from simulated coherence curves. The resulting performance,
	\begin{equation}
		\mathrm{MAE}=0.078,\qquad R^2=-4.05,
	\end{equation}
	demonstrated that the model failed to generalize reliably.
	
	The poor performance is attributable to several factors. First, the training data were generated from simplified analytical coherence functions rather than full quantum-dynamical simulations. Second, a single scalar prediction cannot adequately represent the temporal variability of $H(t)$. Third, the CNN architecture with global pooling suppresses sequential information that is essential for identifying non-stationary noise patterns.
	
	These observations indicate that reliable extraction of time-dependent noise parameters will require sequence-aware architectures such as LSTMs or Transformers, together with significantly larger datasets and physics-informed constraints.
	
\subsection{Limitations and Open Questions}

Several limitations remain in the present framework. The discrepancy between the fitted and theoretical PSD exponents indicates that finite‑time simulations do not fully capture asymptotic low‑frequency scaling; longer sampling and improved spectral estimators~\cite{Thomson1982} may be required. The locally stationary approximation also becomes marginal when $|H'(t)|$ is sufficiently large, potentially demanding the exact Riemann–Liouville kernel~\cite{Ayache2013} for accurate covariance generation.

The Born–Markov treatment~\cite{Breuer2002} employed here neglects strong memory effects that may become important at millikelvin temperatures~\cite{Caldeira1983,Hu1992}. Similarly, the semi‑analytical fallback used after QuTiP convergence failures~\cite{Johansson2013} cannot reproduce the full dynamics of time‑dependent collapse operators, highlighting the need for more robust numerical solvers.

Another open issue concerns the physical interpretation of the effective spectral density. Although the phenomenological bath reproduces experimentally realistic decoherence timescales~\cite{Ithier2005,Bylander2011}, a microscopic derivation from interacting two‑level systems~\cite{Muller2019} or fluctuating defect spins~\cite{Faoro2008,Paladino2014} would provide stronger theoretical grounding. Ultimately, connecting the mmfBm parameters to material‑specific microscopic models remains an important goal for predictive noise engineering.
	\subsection{Future Directions}
	
	Several improvements can substantially strengthen the model. Longer stochastic trajectories generated through circulant embedding~\cite{Dietrich1997} would improve PSD convergence and enable more accurate low‑frequency analysis. Wavelet or multitaper spectral estimation techniques~\cite{Thomson1982} could further reduce bias in non‑stationary noise characterization.
	
	Beyond the Born--Markov approximation, hierarchical equations of motion (HEOM)~\cite{Tanimura2006} or exact non‑Markovian solvers~\cite{Breuer2002} should be implemented to quantify memory effects rigorously. More realistic simulations should also include simultaneous charge and flux noise~\cite{Paladino2014,Krantz2019}, finite pulse durations, and leakage into higher transmon states, which may require optimal control techniques such as DRAG pulses~\cite{Motzoi2009}.
	
	From a machine‑learning perspective, future work should employ Transformer‑based sequence models or physics‑informed neural networks~\cite{Raissi2019} trained directly on master‑equation trajectories. Bayesian approaches~\cite{Granade2012} could additionally provide uncertainty quantification for experimentally inferred noise parameters. Broader machine‑learning strategies for quantum systems are reviewed in Ref.~\cite{Carleo2019}.
	
	Experimentally, the model can be tested through Ramsey and echo spectroscopy on transmon or fluxonium devices over the temperature range $10$--$500~\mathrm{mK}$~\cite{Bylander2011,Ithier2005}. In particular, measurements of stretched‑exponential decay exponents and temperature‑dependent crossover behavior would provide a direct validation pathway for the proposed mmfBm framework.
	
	\section{Conclusion}
	
	In this work, a memory multi-fractional Brownian motion framework was extended to a quantum Caldeira--Leggett environment in order to investigate non-Markovian decoherence in superconducting qubits. The simulations reproduced several physically relevant features, including stretched-exponential Ramsey decay, temperature-dependent crossover behavior, and non-trivial dephasing-rate scaling.
	
	The results further demonstrate that time-dependent long-memory noise can strongly influence both coherence dynamics and gate optimization. At the same time, the simulations exposed important numerical and theoretical limitations, particularly in spectral estimation, ultra-low-temperature integration stability, and machine-learning-based noise extraction.
	
	Overall, the present study establishes a flexible foundation for combining stochastic fractional noise models with open quantum system dynamics. Future developments incorporating exact non-Markovian solvers, improved spectral methods, and experimentally calibrated microscopic baths may provide a more complete understanding of decoherence processes in next-generation superconducting quantum hardware.

	\bibliographystyle{unsrt}
	
	\bibliography{document12}
\end{document}